# High-speed photonic crystal modulator with non-volatile memory via structurally-engineered strain concentration in a piezo-MEMS platform

Y. Henry Wen,[1,*] David Heim,[1] Matthew Zimmermann,[1] Roman A. Shugayev,[3] Mark Dong,[1,2] Andrew J. Leenheer,[3] Gerald Gilbert,[4] Matt Eichenfield,[3] Mikkel Heuck,[2,5,‡] and Dirk R. Englund[2,†]

[1]*The MITRE Corporation, 202 Burlington Road, Bedford, Massachusetts 01730, USA*
[2]*Research Laboratory of Electronics, Massachusetts Institute of Technology, Cambridge, Massachusetts 02139, USA*
[3]*Sandia National Laboratories, P.O. Box 5800 Albuquerque, New Mexico 87185, USA*
[4]*The MITRE Corporation, 200 Forrestal Road, Princeton, New Jersey 08540, USA*
[5]*Department of Electrical and Photonics Engineering, Technical University of Denmark, 2800 Lyngby, Denmark*
[*]*hwen@mitre.org*
[‡]*mheu@dtu.dk*
[†]*englund@mit.edu*

**Summary**

Numerous applications in quantum and classical optics require scalable, high-speed modulators that cover visible-NIR wavelengths with low footprint, drive voltage (V) and power dissipation. A critical figure of merit (FoM) for electro-optic (EO) modulators is the transmission change per voltage, $dT/dV$. Conventional approaches in waveguided modulators seek to maximize $dT/dV$ by the selection of a high EO coefficient or a longer light-material interaction, but is ultimately limited by nonlinear material properties and material losses, respectively. Optical and RF resonances can improve $dT/dV$, but introduce added challenges in terms of speed and spectral tuning, especially for high-Q photonic cavity resonances. Here, we introduce a cavity-based EO modulator to solve both trade-offs: (i) it eliminates the tradeoff between $dT/dV$ and waveguide loss -- i.e., it is possible to reduce $(V L)_\pi$ given a fixed EO coefficient and fixed device length for a single-pass waveguide or fixed-Q cavity; and (ii) it incorporates a non-volatile (NV) cavity tuning $\Delta\nu_{cav,NV}$ for passive memory and programming of multiple devices into resonance despite fabrication variations. Based on this structurally engineered EO modulation, we introduce a photonic crystal cavity (PhCC) modulator based on a silicon nitride (SiN) waveguide-on-aluminum nitride (AlN) actuator platform fabricated in a 200 mm photonic integrated circuit (PIC) platform with transparency across both the visible and infrared spectra. For (i), our approach concentrates the displacement of a piezo-electric actuator of length L and a given piezoelectric coefficient into the PhCC, resulting in $dT/dV \propto L$ *under fixed material loss* -- i.e., the EO FoM can be increased principally indefinitely in the design stage. For (ii), we employ a material deformation that is programmable under a "read-write" protocol with a continuous, repeatable tuning range of 5 ± 0.25 GHz and a maximum non-volatile excursion of 8 GHz. In telecom-band demonstrations, we measure a fundamental mode linewidth, $\gamma = 2\pi \times 5.4$ GHz, with voltage response $\partial_V \nu = 177 \pm 1$ MHz/V corresponding to $\Delta\nu_{cav,DC} = 40 \pm 0.32$ GHz for voltage spanning ±120 V, 3dB-modulation bandwidth $\omega_{BW,3dB}/2\pi = 3.2 \pm 0.07$ MHz broadband DC-AC, and 142 ± 17 MHz for resonant operation near 2.8 GHz operation, optical extinction down to min(log(T)) = -25 dB via Fano-type interference, and an energy consumption down to $\delta U/\Delta\nu_{cav} = 0.17$ nW/GHz. The programmability of these PhCC modulators -- in both the design-specified $\partial_V \nu$ and a field-programmable nonvolatile $\Delta\nu_{cav,NV}$ -- readily extends into the VIS and UV spectrum, as well as to full optical programmability via dual-actuation IQ modulation. More broadly, the strain-enhancement methods presented here are applicable to study and control other strain-sensitive systems.

## Introduction

Numerous applications in quantum and classical optics, including optical neural networks for machine learning, phase-array-based beam control for remote ranging and directed energy, and atom-photon control and interference for quantum information science, motivate the development of scalable, high-speed photonic integrated circuits (PICs) that cover visible-NIR wavelengths [1-11]. Conventional single-pass modulators are bound by the fundamental trade-off between modulation efficiency and material loss and nonlinearity [12-14]. Resonant modulators enhance the modulation efficiency by a factor of the cavity finesse for a fixed waveguide length but are sensitive to spectral inhomogeneity arising from fabrication and environmental variations between device [8,14-16].

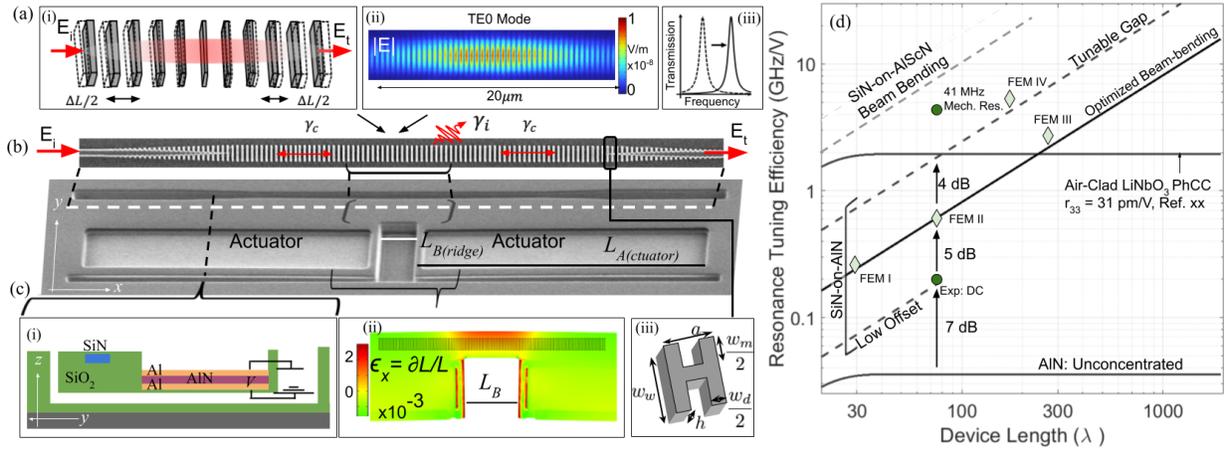

Figure 1: (a) Strain modifies period and duty cycle of photonic crystal waveguide (i) and shifts resonance frequency of cavity modes (ii) to modulate transmission (iii). (b) SEM of SiN waveguide-on-AlN actuator device. The photonic crystal waveguide is embedded within the top strip of the structure connecting the two actuators forming a doubling clamped cantilever. The cavity section of the PhC waveguide is in the narrow bridge where strain is concentrated. (c) Cross-section of released structure showing layer stack composed of Al-AlN-Al piezo actuator connected to SiN waveguide clad in silica (i). The structure is released and free to move within the doubly clamped boundaries and strain concentrated across the bridge (ii) with dimensions shown in (iii). (d) Strain-concentration enhanced EO resonance tuning as a function of device actuator length. Experimental measurements (circles) show 7dB enhancement over devices without strain concentration. Optimization of the strain-optic overlap in FEM (diamonds) gives further 5 dB improvement. Linear increase of the cavity tuning rate with device length can be achieved for device geometries that maintain actuator stiffness and high strain-optic overlap. FEM predicts an additional 4 dB enhancement for a tunable gap device where the two sides of the photonic crystal are mechanically disconnected by a sub-wavelength air gap (see fig. 3). Performance exceeding integrated Lithium Niobate Pockels effect-based photonic crystal modulator [18] is achievable with device length less than 300λ.

Here, we introduce an approach that decouples the length of the electro-optic active region from the size of the electro-optic material and thus enables the magnitude of the electro-optic phase shift to increase without increasing the length of the active waveguide region. This is achieved via a strain-concentration mechanism, shown in Figure 1, in a waveguide-on-actuator piezo-MEMS platform that is configured such that the electro-optically active length of the waveguide is placed in a bridge section of the structure that is more mechanically compliant than the rest of the structure, resulting in a preponderance of the strain from the actuators ($2L_A$) being concentrated within the bridge ($L_B$). This approach is particularly advantageous in a photonic crystal cavity where the optical mode can be localized within the same small region allowing for an optimal overlap of the strain and optical fields.

     We quantify this advantage by considering transmission through a photonic crystal cavity. Figure 1a illustrates a two-sided cavity-modulator in which an input field $E_i$ couples via a reflector to the cavity field, which in turn couples to the transmitted field $E_t$. The voltage-induced change in the on-resonance

cavity transmission $T(\Delta\omega) = \frac{1}{|1+i\Delta\omega/\gamma|^2}$ is given by $\frac{dT}{dV} = \frac{d}{dV}\left[\frac{1}{|1+i\Delta\omega/\gamma|^2}\right] = \frac{-2Q}{|1+i\Delta\omega/\gamma|^3}\frac{d}{dV}\left[\frac{\Delta\omega}{\omega_c}\right]$, where $\gamma = \gamma_i + \gamma_c$ and $\gamma_i$, $\gamma_c$ are intrinsic loss rate and cavity-waveguide coupling rate, respectively, and where $\frac{\Delta\omega}{\omega_c} = -\frac{1}{2}\left[2\Delta n(V)/n + \beta \cdot \epsilon_{Wg,x}(V)\right]$, for uniform axial strain of the waveguide $\epsilon_{Wg,x}(V)$. The first term $(2\Delta n(V)/n)$ is the purely photo-elastic component, that is, the voltage induced change in the dielectric permittivity as seen by the optical mode. The second term $(\beta\epsilon_{Wg,x}(V))$ is moving boundary term where $\epsilon_{Wg,x}(V) = \Delta l_{Wg}(V)/l_{Wg}$ is the voltage induced axial strain on a short segment of the active waveguide $l_{Wg}$ scaled by $\beta = \left[(\varepsilon_1 - \varepsilon_2)|\bar{E}_{\parallel}|^2 + (1/\varepsilon_1 - 1/\varepsilon_2)|\varepsilon\bar{E}_{\perp}|^2\right]/|\varepsilon\bar{E}|^2_{max}$ which arises from a perturbative analysis of moving boundaries [17]. The $\beta$ factor is evaluated at the dielectric boundary of a photonic crystal composed of alternating dielectric elements having relative permittivities $\varepsilon_1 = (n_1)^2$ and $\varepsilon_2 = (n_2)^2$ and represents the conversion factor between strain and the fractional frequency shift. N.B.: 1) $\varepsilon$ is used for relative permittivity and $\epsilon$ is used for strain; 2) $\bar{E}$ is used for electric fields and $E$ is used for Young's Modulus.

In a strain concentrating structure the moving boundary term can be related to the strain of the larger actuator section such that $\frac{\Delta\omega}{\omega_c} = -\frac{1}{2}\left[2\Delta n/n + \beta \cdot C \cdot \epsilon_{A,x}(V)\right]$ where $\epsilon_{A,x}(V) = \Delta L_A(V)/L_A = d_{x3} \cdot V/h_{AlN}$ is the voltage induced strain of the two piezo-actuators each having length $L_A$, an electrode distance equivalent to the piezo-layer thickness $h_{AlN}$, and where $d_{x3} = f(d_{13}, d_{33})$ is a structure-dependent function of the piezo-electric coefficients ($d_{13}$ = 2.5 pm/V, $d_{33}$ = 5 pm/V) effective in the axial (x) direction.

The full description of the strain-optic transfer requires a tensor representation within a finite element mode, which is presented later in detail. For conceptual understanding we consider ideal rigid actuators which only extend or compress axially, in which case the strain concentration factor simplifies to $C = 2L_A/L_B$ which reflects the concentration of strain from the two actuator section of combined length $2L_A$ fully into the shorter bridge section $L_B$. The actuator strain $\epsilon_A(V)$ at a given voltage is a constant with respect to actuator length $L_A$ but $C$ scales linearly with increasing actuator length and decreasing bridge length. This represents the mechanism for decoupling the electro-optic strength from the active waveguide length, and is quantified in figure 1d where the transmission modulation for strain-concentrating devices of increasing actuator lengths is compared with the performance of an air-clad lithium niobate photonic crystal modulator with a fixed length of 15 micron [18]. While the unconcentrated strain from Aluminum Nitride is relatively weak, the concentrated strain from device with combined actuator length of $2L_A$ = 240λ total length $L_{tot}$ = 250λ results in performance similar to integrated Lithium Niobate devices. With improved strain transfer, for example in air-clad or air gap structures shown in figure 3 where the primary resistance to moving waveguide boundaries are removed, a further 4 dB enhancement is predicted, and could be applied to other moving boundary-based mechanisms [19]. This approach is agnostic to the particular piezoelectric material used, thus, integration of stronger piezoelectric materials, such as Aluminum Scandium Nitride ($d_{33}$ = 25 pm/V) [20] and PZT ( $d_{33}$ = 500 pm/V) will proportionally enhance modulation efficiency and reduce actuator footprint. For larger actuators the finite rigidity becomes a limiting factor to effective strain concentration and $C$ becomes a function of the structural geometry and mechanical response of the actuator-bridge structure leading to a multi-parameter space in which to optimize the rigidity and strain concentration. Nevertheless we show via a full electro-opto-mechanical finite element model (FEM) that it is possible to achieve strain concentration of nearly 20 dB within the realistic parameters of the platform for actuators with combined length that is less than 300λ.

Based on this approach we demonstrate a photonic crystal cavity-on-actuator modulator that exhibits DC tuning across 40 GHz with a maximum observed resonance frequency tuning efficiency of 177 MHz/V which represents a 7 dB enhancement upon previous demonstrations in the same platform using unconcentrated strain [10] and similar performance to a large displacement device [11] but with a 100x reduction in actuator foot-print. This work also achieves similar or better resonance tuning efficiency to piezo-PIC platforms based on PZT, a material with a >100× stronger piezoelectric constant [6], but where strain concentration is not leveraged. For AC signals we observe a highly structured mechanical response spectrum which we use to achieve 1) high-fidelity modulation of PRBS-signals from DC up to 3 Mbps via coherent optimal control-based pre-compensation, 2) data rates of 120 Mbps using modulated carrier signals, and 3) resonant modulation using mechanical modes oscillating at frequencies up to 2.8 GHz and, for resonances in the 20-200 MHz range, the resonantly enhanced tuning efficiency reaches 2.2 GHz/V. This mechanically resonant enhancement can be additionally combined with electrically resonant circuits to improve resonant operation even further. While a piezo-based platform requires higher voltages than electro-optic platforms, the power dissipation for our device is sub-100 uW for 1-MHz rate modulation with a holding power of < 1 nW for ±30V, making it compatible with applications that require cryogenic temperatures. At high voltages the device exhibits stable and repeatable hysteresis which we attribute to mechanical buckling-based multi-stability. This feature enables repeatable non-volatile memory in the form of 5 GHz of repeatable non-volatile cavity resonance tuning and maximum non-volatile excursions of up to 8 GHz with long term stability, enabling zero-power static tuning of the optical transmission. The extinction ratio is currently limited by the quality factor of the photonic crystal cavity. For a given quality factor the extinction ratio of a device can be further amplified by combining a PhC cavity modulator with a PhC mirror in a Michelson interferometer configuration to achieve > 25dB extinction ratio. We also demonstrate operation at visible wavelengths on the same wafer wherein the photonic crystal is defined via electron beam lithography, which illustrates the path towards scalable control of quantum systems in the visible range. The combination of fast modulation with non-volatile memory in a small footprint presents a uniquely flexible device for the development of scalable integrated photonic circuits. In the subsequent sections we first present the device design and fabrication, followed by results sections on: 1) DC tuning, 2) FEM and enhanced strain concentration designs, 3) Non-volatile tuning via mechanical multi-stability, and 4) Resonant AC Modulation and Coherent Pre-Compensation-based PRBS Modulation.

**Device Design and Fabrication**

The PhC modulator devices are fabricated in a 200mm-wafer CMOS platform with silicon nitride (SiN) waveguides cladded by silicon dioxide ($SiO_2$) and mechanically coupled to an aluminum nitride (AlN) piezo-electric stack, see Fig. 1 and Refs. [10,11] for further details. Fig. 1(a,b) shows the layout of the modulator device where the PhC cavity is placed on a narrow bridge of length $L_B = 12$ µm and width $W_B = 6$ µm between two opposing piezo-actuators each of length $L_A = 64$ µm and width $W_A = 18$ µm. Removing a sacrificial amorphous silicon layer underneath the piezo stack releases the actuators from the silicon substrate to form a suspended double-clamped cantilever, see Fig. 1(b). The intrinsic stress built into the film layers makes the cantilever bend downward, as shown in Fig. 1(d). Applying a positive voltage across the AlN causes the actuators to expand and deflect vertically and laterally as illustrated in Fig. 1(e). The weak connection formed by the bridge at the center of the cantilever concentrates the actuator-induced strain to maximally overlap with the optical cavity mode. Strain deforms the shape of the PhC cavity which induces both photo-elastic and moving boundary effects, shifting of the resonance frequency of the cavity.

The PhC cavity is formed by periodically modulating the width, $w_w$, of a straight waveguide, see inset in Fig. 1(a). The waveguide is connected to the PhC cavity via taper sections that linearly vary the modulation width from $w_m=0$ to $w_m=w_w$. This complete width-modulation inside the cavity region maximizes the mirror strength of the PhC [22,23] and minimizes the rigidity of the bridge to amplify the strain concentration. The central part of the cavity is formed by gradually modifying the duty cycle, $w_d/a$, such that the mirror strength of each unit cell varies linearly with distance [23]. See Supplementary section 1A for more details and specific parameters. Our standard fabrication process relies on deep-ultraviolet optical lithography, which has sufficient resolution to define PhCs that operate in the telecom range at wavelengths around 1550nm. For operation near 780 nm, the SiN patterning is performed using electron-beam lithography and is left unclad and unreleased due to current fabrication limitations.

**Results: DC Tuning & High-Extinction Ratio Configuration**

The optical properties of the PhC devices are characterized by measuring the transmission of a wavelength-tunable continuous-wave (CW) laser through the device via grating couplers. Modulation of the optical transmission is measured by connecting a ground-signal-ground RF probe to metal contacts at the chip surface that are connected to the piezo-actuators through vias and a routing metal layer [10,11]. Fig. 2 (a) plots the measured optical transmission spectrum of a representative telecom-band PhC device showing two resonances and an extinction inside the photonic bandgap of ~50 dB. Fitting the spectra to a Lorentzian function yields $\lambda_{TE1}$= 1553nm, $Q_{c,TE1} = 1.2 \times 10^5$, and $Q_{i,TE1} = 2.8 \times 10^4$ for the TE1 mode, as well as $\lambda_{TE0}$= 1541 nm, $Q_{c,TE0} = 3.5 \times 10^6$, and $Q_{i,TE0} = 2.3 \times 10^4$ for the TE0 mode. The quality factors $Q_c$ and $Q_i$ correspond to waveguide-coupling and intrinsic loss mechanisms, respectively. Each device was optimized for coupling to one of the two cavity modes within the band-gap. Devices where both cavity modes were discernible were measured for both. Fig. 2(b) shows transmission spectra for a device optimized for the TE0 mode with a 12 µm bridge for different DC voltages ranging from $V$=-30V to $V$=+30V exhibiting a maximum extinction ratio reaching 13.7dB. Since the modulation efficiency $\partial_V T$ is subjective to difference in resonance Q factor between devices, we present and compare the devices using the frequency tuning efficiency $\partial_V \nu = \partial\nu/\partial V$, which is a direct measure of the strain-induced phase shift.

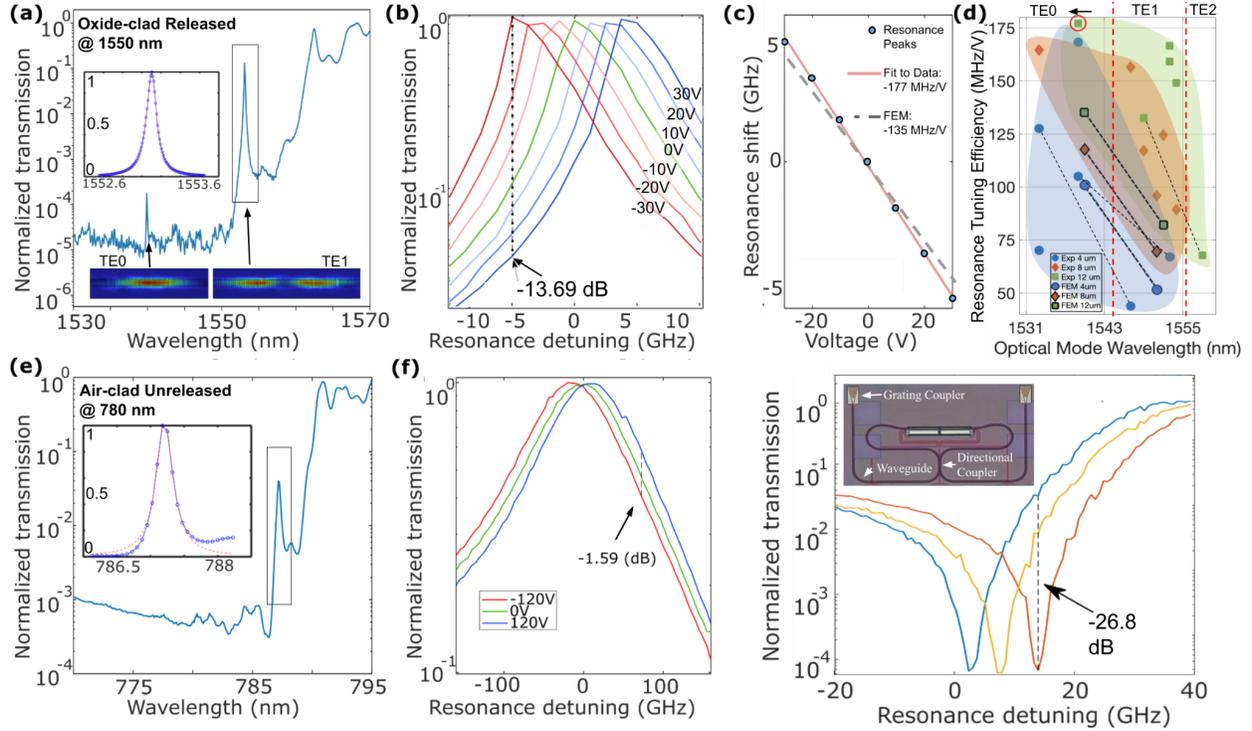

**Figure 2:** (a) Transmission spectrum of typical PhC modulator with an inset showing a Lorentzian fit (red) of the measured resonance (blue) at ~1553nm. Mode shapes of TE0 and TE1 shown for the two resonances. (b) The shift in resonance frequency from applying a DC bias to the piezoelectric actuators of +-30V. (c) Linear fit of the DC voltage sweep in GHz per volt. (d) Tuning efficiency of 15 measured devices as a function of the optical mode order, wavelength and bridge length, overlaid with the FEM results (hard-bordered) for each device type and mode order. Connected points correspond to the same device exhibiting multiple optical modes. The device shown in (a-c) circled in red. (e) Spectrum of a visible-wavelength PhC modulator with a measured resonance at ~787.5nm. f) Strain tuning of an unreleased visible-wavelength PhC modulator. (g) Coupled PhCC devices in Michelson interferometer creates sharp Fano-type resonances with 26.8 dB contrast 0 to 130V.

The DC tuning efficiency is extract by fitting a line to the resonance frequency as a function of DC voltage as shown in Fig. 2(c). Fig 2(d) summarizes the DC tuning of 15 distinct devices with varying bridge lengths and optimized for one of two cavity modes. The spread in resonance wavelength and tuning efficiencies results from the variations in thickness and intrinsic stress of the constituent film layers, which are measured to vary between 5-10% across a single wafer. The extracted DC tuning efficiency is a combination of both in-plane and out-of-plane contributions, a distinction which is discussed in the section on non-volatile tuning. Results from an unreleased but air-clad visible wavelength device are presented in Fig. 2(d,f). The photonic crystal waveguide was fabricated using e-beam lithography after the main photo-lithography process flow. The single resonance at $\lambda_{res}$=787.5nm has $Q_c$= $2.5\times10^4$, $Q_i$= $3.04\times10^3$, and $\partial_V \nu$ = +100 MHz/V which is $10^2\times$ enhanced over the expected efficiency for an equivalent unreleased oxide-clad device based on FEM simulation. The modulation contrast is reduced with respect to the telecom devices due to the lower Q associated with higher absorption and scattering at lower wavelengths. However, this demonstration indicates that further loss engineering and alternative methods for releasing air-clad photonic crystals will enable modulation efficiencies in the 10s of GHz/V.

The extinction ratio is currently limited by the loss due to the sidewall roughness of the waveguide (-2 dB/cm) to which the photonic crystal mode is particularly sensitive. Alternative etch processes have reduced this by a factor of 4, from which we expect an extinction ratio of 25.5 dB for ± 30V. A complimentary method for achieving higher extinction makes use of a Michelson-interferometer (MI) consisting of two one-sided PhC cavities connected via a directional coupler as shown in Fig. 4(a) [24-26]. Interference between the two reflection ports results in sharp spectral features from Fano-like resonances that results in an extinction ratio of 26.8 dB. The MI devices may also be driven in a push-pull configuration like Mach-Zehnder interferometer modulators to reduce the required drive voltage. Note that even though our measured transmission spectra indicated a resonance in only one of the cavities, the transmission spectra retains sharp features as shown in Fig. 4(b). Using temporal coupled mode theory to model the transmission [12] (see Supplementary section 1B for details) and fitting to the measured data, we find a resonance wavelength of 1565.7nm and a quality factor of $Q = \omega/\kappa = 1.1 \times 10^4$. Note that we can only estimate the total $Q$ due to an over-complete set of model parameters. The transmission in Fig. 4 (b) drops by almost 40 dB over a very small frequency range and with a DC bias of 130V we measure an extinction of 26.8dB as shown in Fig. 4 (c). The actuator design for the MI configuration is a single-clamped cantilever and the resulting modulation efficiency is $\partial_V \nu$ = +7.7 MHz/V. These singly clamped devices, originally designed to fit both elements into the same-sized socket, do not have the strain concentration mechanism of the doubly-clamped devices. Implementation of this device using the doubly clamped strain concentration mechanism is expected to provide both enhanced modulation efficiency and extinction ratio.

**Results: FEM Simulations & Enhanced Strain Concentration Designs**

We construct a finite-element model of the released telecom-band device in COMSOL Multiphysics to model the combined electronic-optical-mechanical behavior of the device. This allows us to understand the mechanism by which strain is transferred to the optical waveguide and to estimate the relative contributions from the photo-elastic and moving boundary effects. The FEM model outputs the strain tensor $\bar{\epsilon}$ for the entire structure from which the moving boundary contribution can be directly evaluated. The photo-elastic contribution is given by the strain-induced refractive index shift, for which the *y*-component is dominant for the TE mode:

$$\Delta n_y = -n^2/2(p_{11}\epsilon_y + p_{12}(\epsilon_z + \epsilon_x)), \quad (1)$$

where $p_{11}$ and $p_{12}$ are estimated from the material specific photoelastic coefficients [5,27], and $\epsilon_x$, $\epsilon_y$, and $\epsilon_z$ are the strain along the three coordinate axes for each point in space within a material as indicated in Fig. 1(a,b). With both effects taken into account, the optical eigenfrequencies for the photonic crystal cavities modes are then obtained, with a ratio of the strain to fractional frequency shift of $\beta/2$ = -0.58/strain. The predominant strain term $\epsilon_x$, the axial strain, is calculated to vary by $\partial_V \epsilon_x = 1.55 \times 10^{-6}/V$. For the device in Fig. 2(a-c), the calculated $\partial_V \nu_{mb}$ = -146 MHz/V and $\partial_V \nu_{pe}$ = +11 MHz/V, where the subscripts *mb* and *pe* refer to *moving boundary* and *photo-elastic*, for a net modulation efficiency of $\partial_V \nu$ = 135 MHz/V, shown in fig 2(d) as the hard-bordered green square under the TE0 column. The tuning efficiencies for devices of varying bridge lengths (4, 8 and 12 $\mu m$) and for the two optical modes are overlaid with the experimental data in figure 2(d) where we see that the devices with longer bridge lengths and acting on the smaller of the two optical modes have higher modulation efficiencies both in measurement and simulation due to the better overlap between the strain and optical fields. The discrepancy between the calculated and measured values can likewise be attributed to deviations between the actual and simulated values for refractive indices, Young's Moduli, film thickness, piezo-electric coefficients and intrinsic stress.

Having established the validity of the FEM we extend the model to chart a path to larger strain-concentration devices. While the strain concentration factor $C$ requires FEM to fully capture we find that tanalytical beam models are able to provide close estimates of the FEM results. Combining insights from the FEM and analytical models, we discern two principles for designing devices with higher strain concentration:

1) Within the bridge section the model confirms that the mechanism of strain transfer to the waveguide is via bending of the waveguide bridge. This results in compression and tension on opposing sides of the beam with a neutral axis within the beam (fig. 3a and insets). The strain in a simple uniform beam at a point $\Delta_{wg}$ normal from the neutral axis is given by the flexural strain formula: $\epsilon_{x,Wg}(\Delta_{wg}) = \Delta_{wg} M/EI$ where $M$ is the bending moment of the beam [33]. The flexural rigidity $EI$ is the product of $E$ the Young's modulus of the beam and $I = a^3 b/12$ the area moment of inertia for a beam with cross-sectional dimensions $a \times b$ and bending in the $\hat{a}$ direction. This expression shows that the magnitude of the strain increases linearly away from the neutral axis. In our current devices the waveguide is centered laterally within the bridge and offset by 1.275 nm from the neutral axis which is still sub-optimal since the optical mode overlaps substantially with the neutral axis and areas of opposing strain, as shown in the fig. 3(b, inset). The adjoining table (fig. 3c) shows that increasing the waveguide offset by 1 μm more than doubles the modulation efficiency. Increasing the offset further has limited gains due the cubically increasing flexural rigidity as the bridge width increases.

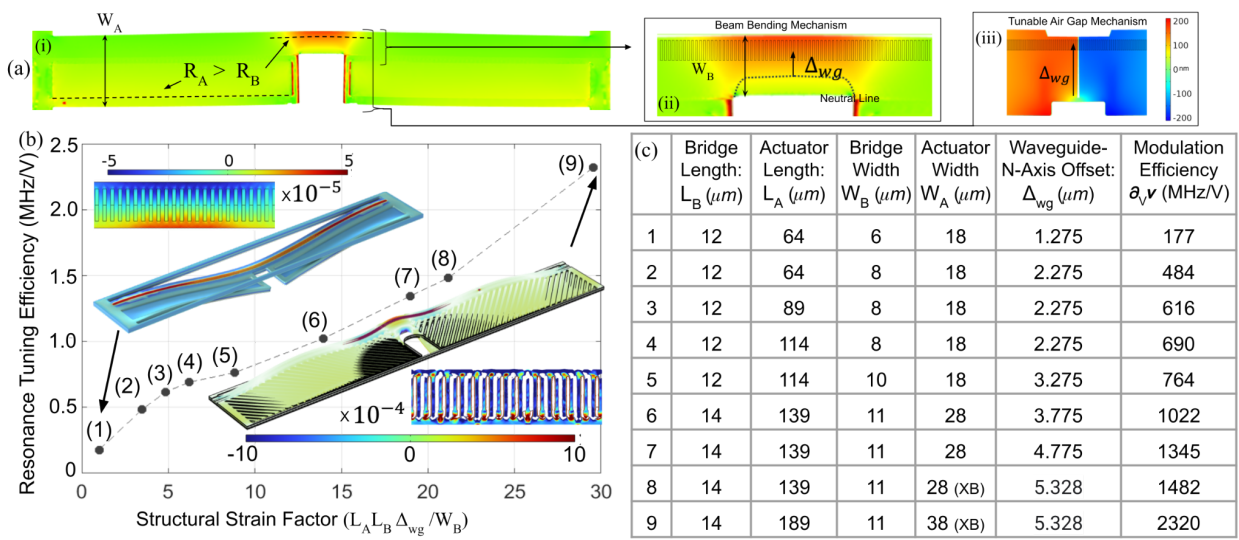

Fig. 3: a) (i) Differential strain distribution across the entire structure with strain concentrated in the bridge. (ii) Bridge is bent with tension above and compression below the neutral line. The difference between the curvatures of the actuators and bridge is highlighted. (iii) Bridge can be replaced with a tunable-gap structure that concentrates the moving boundary effect within a single PhC period. (b) Resonance tuning efficiency of enhanced strain concentration designs as a function of a structural strain factor ($L_A L_B \Delta_{wg}/W_B$) normalized to the value for the experimentally demonstrated structure. It is used as an analytical approximation of the strain concentration factor $C$. $L_B$ is in the numerator since the current bridge is shorter than the cavity mode. Top left inset: FEM model and strain profile at the PhCC of the experimental device. Bottom right inset shows the same for design 9 where the strain in the cavity mode volume is far from the neutral axis. The specific structural parameters are given in the table in (c). All simulated designs in (b,c) are beam-bending devices.

2) Despite the sub-optimal waveguide placement of the current devices we find that the strain concentration is more efficient, by ~18 dB, if directed laterally (in plane) by putting the waveguide bridge off-x-axis to the actuators versus vertically (out-of-plane) but putting them symmetrically on-x-axis. Analytically, this is due to the difference in the relative magnitudes of the in-plane and out-of-plane flexural rigidities of the actuator and bridge sections, $(EI)_{A,y}/(EI)_{B,y} > (EI)_{A,z}/(EI)_{B,z}$. The curvature for a beam $\kappa$ with applied force is inversely proportional to $EI$. For a uniform Young's modulus beam bending out-of-plane, $a = h = 2.2\ \mu m$ for both the actuator and bridge and the ratio of the curvatures $\rho_{\kappa,z} = \kappa_{B,z}/\kappa_{A,z} = W_A/W_B = 3$ indicating that the bridge bends 3× more than the actuators. Laterally the role of height and widths are reversed and $\rho_{\kappa,y} = \kappa_{B,y}/\kappa_{A,y} = (W_A/W_B)^3 = 27$, a 9× enhancement over vertical bending. Since the amount of axial stress that must be balanced is conserved between the two cases, lateral bending concentrates far more of the strain in the bridge section than vertical bending. Since the actuators are not perfectly rigid we find that the modulation efficiency does not scale directly with the 3rd power of the actuator width or linearly with the actuator length. Rather, as the actuator dimensions increase it is more prone to out-of-plane deformation that reduces the effective scaling. This can be partially mitigated via an anisotropic increase of the actuator's flexural rigidity by patterning the top oxide layers to increase its lateral rigidity while not significantly affecting its axial rigidity, for example with diagonal crossbars. Such a structure channels the actuator strain more effectively towards the bridge while not constricting axial motion. Lastly, the length of the bridge section must also be optimally matched to the size of the optical mode so that the strain field and E-field overlap optimally which, based on the measured devices, requires a longer bridge length to match the optical mode size.

Combining these design principles we extend the FEM to simulate a device with two actuators each of $L_A = 140\ \mu m$, $W_A = 28\ \mu m$ and diagonal oxide crossbars for additional out-of-plane rigidity and strain channeling. The bridge has dimensions $L_B = 14\ \mu m$ and $W_B = 10\ \mu m$ and the waveguide displaced $3.5\ \mu m$ laterally from the center of the bridge. The design and performance of this design is presented in Figure 3 and table 1 in comparison to the fabricated design and a range of intermediate variations to demonstrate the effect of individual parameters. Altogether these design changes result in a predicted modulation efficiency of 2.3 GHz/V, which is comparable to the performance of an integrated Lithium Niobate modulator of similar length and similar footprint [18].

**Results: Non-volatile Tuning via Mechanical Multi-stability**

Non-volatile tuning is a highly desired feature for memory and trimming purposes [16, 29-32]. For DC voltages above a threshold value in the range of ± 20-40 V the devices exhibit hysteresis where the resonance wavelength back at 0 V is shifted compared to its prior value and increases monotonically with the applied write voltage ($V_{\text{write}}$). The hysteretic tuning loops for three distinct devices are presented in Fig 5(a) by cycling the voltage as follows: $V = 0V \rightarrow V_{\text{write}} \rightarrow 0V \rightarrow -V_{\text{write}} \rightarrow 0V$. This is performed 10 times in 5 volt increments each for $V_{\text{write}} = 40\ V$, 80 V and 120 or 125 V. For the F1 device the 20 V loop is included since the hysteresis threshold is below 40 V. The transmission spectrum at $V = 0V$, depends on whether the previous extremal voltage was at $-V_{\text{write}}$ or $+V_{\text{write}}$. The two branches of the loop overlap for the lowest voltage loop ($V_{\text{write}} = 40\ V$ for the F0 and F3 devices, and $V_{\text{write}} = 20\ V$) but clearly separate for higher voltage loops, resulting in an effective tuning efficiency that is different from that of the linear tuning range. When the device is unloaded, that is, at the maximum voltage of a particular loop when the applied stress is reversed the device again exhibits a linear tuning efficiency and total span close to that of the original linear range centered at 0 V, but with a static offset -- in both center voltage and center wavelength -- that depends on the previous maximum voltage. These can be conceptually visualized as a semi-continuous series of stable stress-wavelength curves as shown in fig 5(d) wherein stress is voltage-controlled and wavelength is strain-controlled. From this we see that F1 and F0 correspond to a yielding nonlinearity wherein the effective Young's modulus decreases, resulting in a net tuning efficiency that is higher than the linear tuning efficiency and a CCW hysteresis loop. Conversely, F3 has a stiffening nonlinearity resulting in lower net hysteretic tuning efficiency and a CW hysteresis loop. Notably, the linear tuning efficiency of all 3 devices have a similar value near 87 MHz/V and a non-volatile tuning efficiency range from 18-20 MHz/V, however the net DC tuning efficiency across the entire ±120V range differ substantially due to the specific arrangement of the non-volatile states. The potential causes for these differences are discussed later in this section.

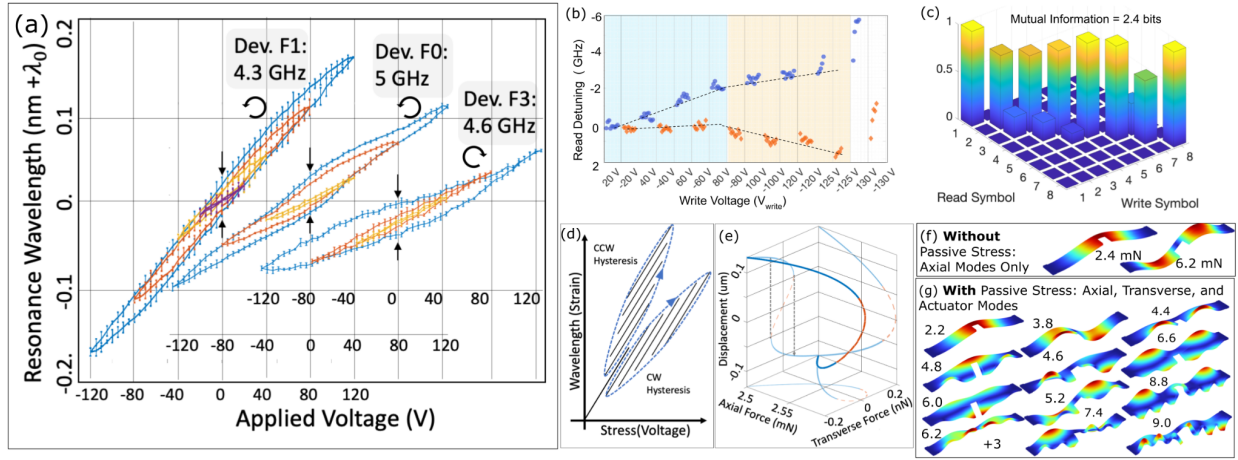

Figure 4. (a) Resonance tuning response for cycling voltage loops of increasing range ($V_{write}$, the "write" voltage) for 3 devices: ± 20V (F1 only) ±40 V - yellow, ±80 V - red, and ±120 (F1) or ±125 V (F0, F3) - blue, each repeated 10 times in 5 volt increments [SI]. Error bars show standard deviation of the 10 repetitions. (b) Non-volatile "read detuning" $\Delta\nu_{read}$ vs $V_{write}$. The voltage is cycled 10 times for each positive and negative value of $V_{write}$, i.e. 10×(0V →20V→ 0V) followed by 10×(0V → -20V→ 0V) without feedback or correction. Data points within each bin are spaced laterally for clarity and correspond to subsequent applications of the same $V_{write}$. (c) Confusion matrix of write-read protocol shown in (b) using only repeatable range (±120V) showing mutual information of 2.4 bits using 8 bin optimized encoding. (d-g) Model for achieving multi-stable NV tuning via beam buckling modes. (d) CCW (CW) hysteresis indicates transition to buckling state with a higher (lower) 0 V wavelength value, corresponding to (F1, F0) and (F3), respectively. (e) Calculation of critical load of first buckling mode based on ref. [28] for uniform beam with dimensions of the bridge (5.9μm×2μm) and the actuator (18μm×2μm) have critical loads at 0.645mN and 2.55 mN (shown), respectively. 2D projections of the stress-displacement curve are shown on the left, right and bottom panels. (e,f) FEM buckling modes (e) without and (f) with intrinsic stress, labeled by its critical load force.

Figure 5(b) shows the NV tuning for multiple positive unipolar cycles, i.e. 0V → $V_{write}$ → 0V, followed by an equal number of negative cycles $V_{write}$ from 20 V to 130 V showing that the NV tuning does not depend on the repetition or duration of the write voltage up to ±120V. Beyond ±120V more dramatic modification is observed for each additional voltage cycle, marking a departure from the repeatable hysteresis of lower voltages. A maximum NV tuning of nearly 8 GHz is obtained for voltages of ±130V. As a read-write NV memory the device must be operated within the repeatable range, i.e. below 120 V. Based on the blind-write measurements show in fig. 5(b) the NV tuning is able to hold a mutual information of 2.4 bits based on an optimized 8-bin encoding, resulting in the confusion matrix encoding show in fig. 5(c). For trimming of multiple cavities into resonance, the fabrication variation-induced detunings of the cavities must be within the NV tuning range. The lack of replica devices within the same die and intentional wafer-to-wafer variations reduced the number of comparable devices. In the best case, two "F0" devices from separate dies of the same wafer had a detuning of 6 GHz. The long-range wafer-scale variations in film thickness and stress values predict a fabrication-induced variation of 13 GHz in FEM. However, die-level variations are measured to be at least 10 times lower than wafer-scale variations, thus we expect replicas devices on the same die to have a large percentage within the 5 GHz range. Finally, to investigate the long term repeatability and stability of the NV tuning we cycle between the two set points $N_{cyc}$ times, once per minute, and record the spectra. Fig. 5(d) plots the results for $V_{max}$ = 120V and $N_{cyc}$= 1400 show that the setting is repeatable with a standard deviation of 250 MHz of the center frequency without additional correction [SI].

We attribute the NV tuning to the mechanical buckling nonlinearity which arises in mechanical beams with sufficiently strong axial compression force ($F_x$). This attribution is supported by white-light interferometry-based vertical displacement measurements which reveals a residual mechanical deformation of the actuators that is proportional to the preceding maximum voltage (SI). Fig 5(d) shows an analytical beam buckling model for a simple uniform beam wherein the transverse displacement becomes perturbatively unstable at the critical load in the neutral configuration and bifurcates into two statically displaced buckling modes [28,33]. The controlling parameters for this model are the length and the flexural rigidity, $EI$, of the beam. Holding the device length (140 μm) and thickness (2 μm) constant and varying the beam width between actuator (18 μm) and bridge (5.9 μm) widths, the model predicts the first buckling mode to have a critical load between 0.645 mN and 2.57 mN, which is within the accessible range of axial force. The threshold voltage for the onset of buckling is estimated to be 5-50V based on a the critical load values of the linear buckling modes predicted in the FEM simulation. In our devices the axial compression comes from the combination of 1) the intrinsic stress of the device material layers, and 2) the piezomechanical stress. All stress is retained within the structure upon release due to the doubly-clamped boundaries. The intrinsic axial stress is estimated to be of $F_x$ = 4 - 5 mN, which is sufficient to place the structure in the regime of higher order buckling modes. Using $d_{33}$= 5 pm/V the piezo-induced axial stress is estimated to be $\partial_V F_x$ = 30-50 μN/V, equivalent to an effective range of $F_x$ from 7.2 to 12 mN for ±120V. This allows the device to access the majority of the buckling modes between 0 and 10 mN.

The critical load values and shapes of the buckling modes within this axial stress range are obtained using the FEM-based linear buckling eigensolver in COMSOL Multiphysics and presented in Fig 5 (e,f). Notably, the intrinsic material stress plays a prominent role for increasing the density of buckling modes within the accessible stress range. A device without intrinsic stress has only 2 predicted buckling modes, which correspond to the simple "lengthwise" beam buckling modes (5e). For a device with intrinsic stress, there arises buckling modes along the width and separately lengthwise within each actuator section, while at the same time reducing the critical load for the higher order full-length-wise modes (5f). Every axially stressed degree-of-freedom can, in principle, support a distinct family of buckling modes. This results in a large number of modes within the accessible range with critical load values close to one another with the separation between any two adjacent modes ranging from 0.2 to 2.2 mN. Using these values and a piezo-stress coefficient of 0.04 mN/V we estimate the lower and upper bound of the buckling threshold to be in the range 5-50 V. Also, the proximity of the critical load of the first FEM buckling mode (2.2 mN) to the upper-bound of the analytical model (2.55mN) is commensurate with the actuators taking up the majority of the device length.

We further note that all of the modes within the accessible stress range are out-of-plane buckling modes. As discussed earlier, these have a lower strain-optic overlap than the in-plane degree-of-freedom on which the majority DC tuning mechanism relies. The first in-plane buckling mode appears at 50 mN, well beyond the current piezo-accessible range, a result of the large difference between the in-plane and out-of-plane flexural rigidities of the entire structure: $(EI)_y > (EI)_z$. The lateral asymmetry of the device also biases the structure away from the lateral neutral state and into a regime where the density of stable states is lower, making the device predominantly linear for in-plane motion and nonlinear for out-of-plane motion. The ratio between the linear tuning efficiency (~87 MHz/V) to the non-volatile tuning efficiency (~19MHz/V) and the resulting shape of the hysteresis loops can be associated with the ratio of the analytical estimates of the strain transduction rates for in-plane and out-of-plane motion: $[\rho_{\kappa,y}\Delta_{wg,y}/(EI)_{B,y}]/[\rho_{\kappa,z}\Delta_{wg,z}/(EI)_{A,z}] \sim 5$. The different orientations of the hysteresis loops is likely a result of whether the device tensions or compresses for increasing voltage which depends on whether the device is passively above or below the neutral plane of the chip. These details are the subject of ongoing investigation.

**Results: Resonant AC Modulation and Coherent Equalization-based PRBS and Pulse Modulation**

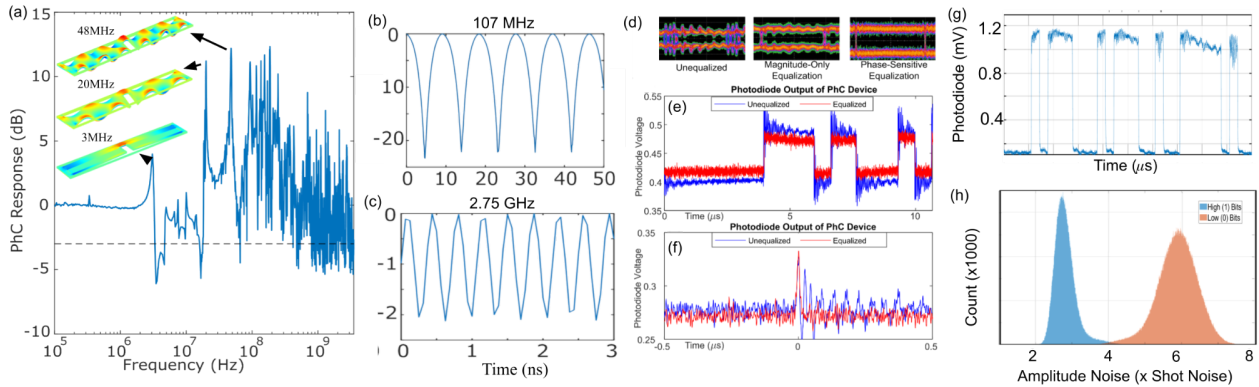

**Figure** 5 (a) The small-signal modulation response ($S_{21}$) for the PhCC modulator from 100kHz up to 3.5GHz, revealing 10-12 dB enhancement at mechanical resonances up to 200 MHz and additional resonances up to 2.8 GHz with 5-10 dB enhancement. Displacement profiles of several resonant modes at 3 MHz, 20 MHz and 40 MHz are shown. (b,c) The resonant response at 107.5 MHz ( 23 dB contrast) and 2.75 GHz (2 dB contrast) for 40-Vpp sine wave. (d-g) Coherent equalization for high-fidelity PRBS and pulse modulation. (d) Eye diagrams of 3 Mbps PRBS-12 signal for (i) an unequalized drive and for equalization using (ii) magnitude-only and (iii) phase-sensitive equalization. (e,f) Comparison of raw time-trace between unequalized and phase-sensitive equalized (e) PRBS-12 and (f) 5 ns gaussian pulse. (g) PRBS-12 data for 80$V_{pp}$ achieving 9 dB contrast. (d) Amplitude noise distribution of all time-bins of the high contrast PRBS-12 signal with pre-compensation taken over 140 repetitions of the PRBS pattern, sampled at 1 GS/s. The high-bits have lower noise with respect to shot noise due a combination of the N scaling and the lower voltage sensitivity near the resonance peak, reaching a value in the range of 2-3x shot noise.

Having established the DC and NV tuning of the device we proceed to demonstrate that it is also capable of fast AC modulation. Here, the resonant mechanical response plays a significant role in providing further strain enhancement on top of the geometric strain concentration. This greatly enhances the resonant tuning efficiency but presents a challenge for the modulation of arbitrary signals. To measure the frequency response, we set the laser wavelength on the blue-detuned slope of the cavity resonance and apply a 6.3V peak-to-peak sinusoidal signal. The optical output is measured with a DC-9.5 GHz photodiode connected to a high-speed sampling oscilloscope (16GS/s @ 4 GHz bandwidth) where the peak-to-peak amplitude of the modulation signal is plotted as a function of the drive frequency. We observe a flat response from DC to 3MHz where a first mechanical resonance is observed, beyond which the response spectrum becomes highly structured with mechanical resonances having up to 15dB enhancement above the low-frequency limit. The optomechanical spectrum continues to be discrete but densely filled up to 500 MHz above which the strength and density of mechanical resonances decreases. The insets show shape and strain profiles of film bulk acoustic resonance (FBAR) modes calculated using COMSOL with eigenfrequencies that agree well with the measured spectrum. Fig. 4(b, c) captures a time domain trace showing >20dB modulation contrast for a 40$V_{pp}$ sine-wave drive at a 107.5MHz mechanical resonance and > 3dB contrast at 2.811 GHz using 6.3$V_{pp}$. This demonstrates both nanosecond-scale rise and fall times and a significant increase in tuning efficiency compared to DC operation. The resonant enhancement can also be leveraged for arbitrary data transmission via a modulated signal centered at a mechanical resonance. Using a QPSK-based modulation scheme we achieved 120Mbps modulation using the mechanical resonance centered at 2.73 GHz (SI).

  Modulation of non-periodic and arbitrary waveforms is complicated by the highly structured frequency response. For example, direct modulation using a square-wave PRBS bit-pattern easily excites the strong mechanical resonances and results in ringing artifacts which overwhelm the output waveform. This is not sufficient for quantum control of atomic systems which require repeatable, high-fidelity pulses with well controlled pulse shape and pulse energy. A well-established technique in telecommunications modulation is the use of pre-compensated waveforms to correct for the uneven, dispersive, or nonlinear transfer function of a particular modulator [34], which works well for free-carrier and Pockels-effect based modulators where the non-flatness of the transfer function is shallow across a large bandwidth. Pre-compensation of sharply resonant modulators is more challenging because of rapid changes in the frequency and phase response over narrow frequency ranges. We demonstrate the ability to compensate for these resonances with the following scheme. First, the time domain signal is generated. We oversample the signal to be able to compensate over a high bandwidth. Next, we convert the signal to the frequency domain using a Fast Fourier Transform (FFT). The FFT bins are then scaled inversely using the frequency response measured with the vector network analyzer. The scaled frequency data is converted back to the time domain with an Inverse FFT (IFFT), and a low-pass filter is applied to reduce aliasing. More details are provided in the [SI].

  We test our equalization scheme on our device by applying an unequalized and equalized signal and observing the optical output using a photodiode with a frequency range from 0-125 MHz and a high-speed oscilloscope. A lowpass filter is applied to both waveforms to avoid aliasing. On both the PRBS and Gaussian pulses, we observe an improved rise time and a significant decrease in ringing and noise after applying the compensation scheme. The 95% settling time is less than 15 ns, which could be improved with a higher bandwidth arbitrary waveform generator and photodiode. The ring-down of the Gaussian pulse, which had been 400 ns on the unequalized Gaussian pulse, has been eliminated. Applying this scheme to high-voltage, high-contrast PRBS modulation, a PRBS-12 bit pattern with 9 dB contrast was achieved with modulation voltage of 80 Vpp shown in fig. 4c. The reduced resonant distortion also reduces the amplitude noise of the modulator to near the shot noise limit. The amplitude noise for each point in the PRBS sequence is calculated using 140 repetitions of the PRBS bit-pattern to calculate the amplitude noise at each point when sampled at 1GS/s. The noise values are then separated between high-bits and low-bits and overlaid as histograms showing noise levels of 2-3× shot noise for the high-bits compared to 10-15× shot noise for un-compensated signals.

**Discussion**

The results here have been presented in the context of a waveguide photonic crystal cavity, however the methods employed -- structurally-engineered strain-concentration, mechanical resonant enhancement, and non-volatile tuning -- are applicable to a wide variety of strain-sensitive systems. For color centers in diamond nano-waveguides, strain-tuning efficiencies have been measured in the range of 0.5-2 PHz/strain, axially for the ZPL frequency and transversely for the ground state splitting [35-38]. Strain concentration into diamond color centers is achievable in the structure presented here by replacing the photonic crystal waveguide with a socket into which a diamond nano-waveguide can be loaded and locked. For the strain levels demonstrated here (1.55e-6/V) a tuning rate of 0.75-1.55 GHz/V can be expected for ZPL tuning of SnV in diamond nano-waveguides, compared to 250 MHz/V without strain concentration [35]. Despite the larger Young's Modulus of diamond (1220 GPa), the smaller cross-sectional dimensions of the diamond waveguide (h = 250 nm, w = 300 nm) has lower flexural rigidity within high-strain region of the structure, leading to a higher resultant strain. Large static strains have been used to substantially increase the coherence times of various color centers, but also result in larger strain distributions which require large tunability to spectrally align [39]. The enhanced strain-concentration designs shown in fig. 3 have the potential for greater than 10× increase in the total concentrated stress, inclusive of both the piezo-induced and intrinsic stresses, resulting in tunable strains in the range of 1-2%, which has the potential to increase the ground state splitting of various color centers to the 10 THz-level. More broadly, the large tunable strains achievable in this cryo-compatible platform may open a path to investigate other strain-sensitive systems such as superconducting materials at the nanoscale[40].

      The non-volatile tuning can be further engineered by modifying the length, shape and stiffness of the waveguide-on-actuator structure. The maximum non-volatile tuning is determined by the range of accessible axial stress. In order to take advantage of the lateral buckling modes the device needs to be symmetric -- to access both polarities of the buckled mode -- and more slender, that is, having a larger length to cross-section ratio which increases both the range of axial stress and the density of buckling modes per unit of axial stress [33]. Since different buckling mode families (vertical, lateral, actuator) can be superimposed, the difference in critical load values between the various buckling mode families forms a natural scaffold on which to tailor the density and discreteness of the final NV density of states. For both memory and resonance trimming applications, the precision of the write process can be improved through repetition of a write-read-adjust process and by use of a small hold voltage to maintain precise alignment. While a hold voltage is not strictly non-volatile, the NV tuning reduces the hold voltages needed to encode values within a 5 GHz span to 1-5V rather than the 25-50V that would otherwise be needed. Lastly, other mechanisms, namely 1) ferroelectricity of the AlN film ($E_c$ = 6 MV/cm) and 2) plasticity of the metal layers ($\sigma_y$ = 124 MPa) can also exhibit mechanical hysteresis but require voltage near 270 V and 110 V, respectively, to access. While these higher threshold voltages make their contribution unlikely, they cannot be fully ruled out since fabrication variations and material imperfections may increase the local strain or reduce the threshold stress, further increasing the number and density of stable states. Future work may in fact seek to include these effects in an effort to achieve

larger and more stable non-volatile tuning.

Lastly, the spectral response of the device can be structurally engineered. The current design is composed of distinct rectangular segments exhibiting discrete resonances. The mechanical resonance scales with the actuator length as $L_A^{-(1 \to 2)}$ with the exact value of the exponent depending on the specific shape of the mode. For example, the first peak at 3 MHz is predicted to upshift in frequency by a factor of 2.7 when the cantilever length is reduced by half with a tradeoff on the DC tuning. While shortening the actuator length presents a trade-off of the DC tuning, the shape of the mechanical resonances can also be structurally engineered to increase the strain overlap with the optical mode to increase speed while not sacrificing the tuning efficiency. The actuator bodies can also be varied in width, mass and rigidity via additional patterning techniques to create a more broadly resonant structure, a technique that has been employed in piezo-based energy harvesting systems [41]. Spectral engineering of both the structure and of the drive signal, as demonstrated using coherent pre-compensation, provides a path to achieve high-fidelity and high-speed modulation in this platform. Given the large and varied parameter space available for engineering of all aspects of the device: DC, AC and NV tuning, we anticipate inverse design and machine learning-based approaches to be particularly impactful in further optimization.

**Conclusion**

We have presented an approach to EO modulation that overcomes the tradeoff between waveguide material length and EO modulation efficiency combined with resonantly enhanced modulation and non-volatile tuning in a single device. This approach is based on a structurally engineered waveguide-on-actuator structure that simultaneously enables strain concentration, supports strong mechanical resonances and exhibits repeatable mechanical hysteresis in an auxiliary degree-of-freedom. We have demonstrated DC tuning efficiency of 177 MHz/V -- a 7 dB enhancement of the tuning efficiency over devices without strain concentration -- with tuning across 40 GHz, and a design path for achieving an additional 10 dB enhancement, reaching an effective EO coefficient that is comparable to that of lithium niobate-based modulators. We have demonstrated resonant modulation with efficiencies up to 4 GHz/V, data-rate up to 142 Mbps using broadband resonances near 2.8 GHz and, by using coherent pre-compensation, high contrast DC-band PRBS modulation near shot noise level. Furthermore, by leveraging the distinct mechanical response of the out-of-plane motion we demonstrate repeatable non-volatile tuning of 5 GHz based on a semi-continuous series of mechanical buckling modes. Further engineering of the buckling state-space and improvements in the write and hold process are promising for substantially increasing the mutual information of the device for use as a non-volatile memory and both range and resolution for spectrally alignment of cavities and emitters. Taken as a whole, this work paves the way for scalable engineering of piezo-MEMS based integrated photonics and charts a path towards investigation of high-strain physics in an integrated, cryogenically compatible platform.

| **Metric Summary** | Unit | Value |
| --- | --- | --- |
| Resonance frequency shift response, dv/dV | MHz/V | 177±1 |
| Tuning range $\Delta v_{cav,DC}$ | GHz | 40±0.32 |
| Non-volatile (NV) tuning range $\Delta v_{cav,NV}$ | GHz | 5±0.25 |
| Modulation bandwidth $\omega_{BW}$ at 3 dB/$2\pi$ for broadband DC-AC | MHz | 3.2 ±0.07 |
| Modulation bandwidth $\omega_{BW}$ at 3 dB/$2\pi$ for resonant operation at 2.7 GHz-2.8 GHz | MHz | 142 ±17 |
| Extinction ratio min(-log(T)) for DC, Fano-resonance, AC @ 107 MHz, & PRBS-12 @ 3Mbps | dB | >13, >25, >22, >9 |
| Operation wavelength, λ | nm | 1520-1560, extensible to down to ~600 |
| Fundamental-mode linewidth γ (GHz) | GHz | 5.4 |
| Energy consumption $\delta U/\Delta v_{cav}$ | nW/GHz | 0.17 |

## Author Contributions

YHW & DH performed the DC and AC measurements, YHW & MZ performed the pre-compensation & PRBS measurements; YHW & RS performed the analytical buckling analysis; YHW, MH & DE developed the concept and model for strain concentration; YHW performed the non-volatile memory measurements, FEM & buckling simulations. DH & MD built the chip-testing apparatus. MH designed the devices. AJL fabricated the devices with supervision from ME. ME, GG, MH & DE supervised the project. YHW wrote the manuscript with support from DH and DE and with input from all authors. YHW acknowledges Genevieve Clark for insightful discussions on strain tuning of diamond color centers.

## Acknowledgements

Major funding for this work is provided by MITRE for the Quantum Moonshot Program. D.E. acknowledges partial support from Brookhaven National Laboratory, which is supported by the U.S. Department of Energy, Office of Basic Energy Sciences, under Contract No. DE-SC0012704 and the NSF RAISE TAQS program. M.E. performed this work, in part, with funding from the Center for Integrated Nanotechnologies, an Office of Science User Facility operated for the U.S. Department of Energy Office of Science. M.H. acknowledges partial support from the Villum Foundation program QNET-NODES, grant no. 37417.

## References

1. W. Bogaerts, D. Pérez, J. Capmany, D. A. B. Miller, J. Poon, D. Englund, F. Morichetti, and A. Melloni, "Programmable photonic circuits," Nature **586**(7828), 207–216 (2020).
2. Taballione, C., Anguita, M.C., de Goede, M., Venderbosch, P., Kassenberg, B., Snijders, H., Kannan, N., Vleeshouwers, W.L., Smith, D., Epping, J.P. and van der Meer, R., 2023. 20-mode universal quantum photonic processor. Quantum, 7, p.1071.
3. Chauhan, N., Isichenko, A., Liu, K. et al. Visible light photonic integrated Brillouin laser. Nat Commun 12, 4685 (2021).
4. Corato-Zanarella, M., Gil-Molina, A., Ji, X. et al. Widely tunable and narrow-linewidth chip-scale lasers from near-ultraviolet to near-infrared wavelengths. Nat. Photon. 17, 157–164 (2023).
5. Warren Jin, Ronald G. Polcawich, Paul A. Morton, and John E. Bowers, "Piezoelectrically tuned silicon nitride ring resonator," Opt. Express 26, 3174-3187 (2018)
6. Lihachev, G., Riemensberger, J., Weng, W. et al. Low-noise frequency-agile photonic integrated lasers for coherent ranging. Nat Commun 13, 3522 (2022).
7. Tian, H., Liu, J., Dong, B. et al. Hybrid integrated photonics using bulk acoustic resonators. Nat Commun 11, 3073 (2020).

8. Menssen, A.J., Hermans, A., Christen, I., Propson, T., Li, C., Leenheer, A.J., Zimmermann, M., Dong, M., Larocque, H., Raniwala, H. and Gilbert, G., 2022. Scalable photonic integrated circuits for programmable control of atomic systems. arXiv preprint arXiv:2210.03100.
9. P. R. Stanfield, A. J. Leenheer, C. P. Michael, R. Sims, and M. Eichenfield, "CMOS-compatible, piezo-optomechanically tunable photonics for visible wavelengths and cryogenic temperatures," Opt. Express **27**(20), 28588–28605 (2019).
10. M. Dong, G. Clark, A. J. Leenheer, M. Zimmermann, D. Dominguez, A. J. Menssen, D. Heim, G. Gilbert, D. Englund, and M. Eichenfield, "High-speed programmable photonic circuits in a cryogenically compatible, visible–near-infrared 200 mm CMOS architecture," Nat. Photonics **16**(1), 59–65 (2021).
11. M. Dong, D. Heim, A. Witte, G. Clark, A. J. Leenheer, D. Dominguez, M. Zimmermann, Y. H. Wen, G. Gilbert, D. Englund, and M. Eichenfield, "Piezo-optomechanical cantilever modulators for VLSI visible photonics," APL Photonics **7**(5), 051304 (2022).
12. Mian Zhang, Cheng Wang, Prashanta Kharel, Di Zhu, and Marko Lončar, "Integrated lithium niobate electro-optic modulators: when performance meets scalability," Optica 8, 652-667 (2021)
13. F. Eltes et al., "A BaTiO3-Based Electro-Optic Pockels Modulator Monolithically Integrated on an Advanced Silicon Photonics Platform," in Journal of Lightwave Technology, vol. 37, no. 5, pp. 1456-1462, (2019)
14. Wentao Jiang, Rishi N. Patel, Felix M. Mayor, Timothy P. McKenna, Patricio Arrangoiz-Arriola, Christopher J. Sarabalis, Jeremy D. Witmer, Raphaël Van Laer, and Amir H. Safavi-Naeini, "Lithium niobate piezo-optomechanical crystals," Optica 6, 845-853 (2019)
15. Linran Fan, Xiankai Sun, Chi Xiong, Carsten Schuck, Hong X. Tang; Aluminum nitride piezo-acousto-photonic crystal nanocavity with high quality factors. *Appl. Phys. Lett.* 15 April 2013; 102 (15): 153507.
16. Youngblood, N., Ríos Ocampo, C.A., Pernice, W.H.P. et al. Integrated optical memristors. Nat. Photon. 17, 561–572 (2023).
17. Johnson, S.G., Ibanescu, M., Skorobogatiy, M.A., Weisberg, O., Joannopoulos, J.D. and Fink, Y., 2002. Perturbation theory for Maxwell's equations with shifting material boundaries. *Physical review E*, *65*(6), p.066611.
18. Li, M., Ling, J., He, Y. *et al.* Lithium niobate photonic-crystal electro-optic modulator. *Nat Commun* 11, 4123 (2020).
19. Jiang, W., Mayor, F.M., Patel, R.N. et al. Nanobenders as efficient piezoelectric actuators for widely tunable nanophotonics at CMOS-level voltages. Commun Phys 3, 156 (2020).
20. L. Hackett, M. Miller, R. Beaucejour, C. M. Nordquist, J. C. Taylor, S. Santillan, R. H. Olsson, M. Eichenfield; Aluminum scandium nitride films for piezoelectric transduction into silicon at gigahertz frequencies. *Appl. Phys. Lett.* 14 August 2023; 123 (7): 073502.
21. C. W. Wong, P. T. Rakich, S. G. Johnson, M. Qi, H. I. Smith, E. P. Ippen, L. C. Kimerling, Y. Jeon, G. Barbastathis, and S.-G. Kim, "Strain-tunable silicon photonic band gap microcavities in optical waveguides," Appl. Phys. Lett. **84**(8), 1242–1244 (2004).
22. Q. Quan, P. B. Deotare, and M. Loncar, "Photonic crystal nanobeam cavity strongly coupled to the feeding waveguide," Appl. Phys. Lett. **96**(20), 203102 (2010).
23. Qimin Quan and Marko Loncar, "Deterministic design of wavelength scale, ultra-high Q photonic crystal nanobeam cavities," Opt. Express 19, 18529-18542 (2011)
24. M. Heuck, P. T. Kristensen, Y. Elesin, and J. Mørk, "Improved switching using Fano resonances in photonic crystal structures," Opt. Lett. **38**(14), 2466–2468 (2013).
25. J. Zhang, X. Leroux, E. Durán-Valdeiglesias, C. Alonso-Ramos, D. Marris-Morini, L. Vivien, S. He, and E. Cassan, "Generating Fano Resonances in a Single-Waveguide Silicon Nanobeam Cavity for Efficient Electro-Optical Modulation," ACS Photonics **5**(11), 4229–4237 (2018).
26. H. A. Haus, *Waves and Fields in Optoelectronics* (Prentice-Hall, 1984).
27. Gyger, F., Liu, J., Yang, F., He, J., Raja, A.S., Wang, R.N., Bhave, S.A., Kippenberg, T.J. and Thévenaz, L., 2020. Observation of stimulated Brillouin scattering in silicon nitride integrated waveguides. Physical review letters, 124(1), p.013902.
28. Mattias Vangbo, An analytical analysis of a compressed bistable buckled beam, Sensors and Actuators A: Physical, Volume 69, Issue 3, 1998, Pages 212-216.
29. Z. Fang, J. Zheng, A. Saxena, J. Whitehead, Y. Chen, and A. Majumdar, "Non‑volatile reconfigurable integrated photonics enabled by broadband low‑loss phase change material," Adv. Opt. Mater. **9**(9), 2002049 (2021).
30. J. Geler-Kremer, F. Eltes, P. Stark, D. Stark, D. Caimi, H. Siegwart, B. Jan Offrein, J. Fompeyrine, and S. Abel, "A ferroelectric multilevel non-volatile photonic phase shifter," Nat. Photonics **16**(7), 491–497 (2022).
31. J.-F. Song, X.-S. Luo, A. E.-J. Lim, C. Li, Q. Fang, T.-Y. Liow, L.-X. Jia, X.-G. Tu, Y. Huang, H.-F. Zhou, and G.-Q. Lo, "Integrated photonics with programmable non-volatile memory," Sci. Rep. **6**, 22616 (2016).
32. C. Ríos, M. Stegmaier, P. Hosseini, D. Wang, T. Scherer, C. D. Wright, H. Bhaskaran, and W. H. P. Pernice, "Integrated all-photonic non-volatile multi-level memory," Nat. Photonics **9**(11), 725–732 (2015).
33. Timoshenko, S. and Gere, J. (1963) Theory of Elastic Stability. 2nd Edition, McGraw-Hill, New York.
34. B. Murray, C. Antony, G. Talli and P. D. Townsend, "Predistortion for High-Speed Lumped Silicon Photonic Mach-Zehnder Modulators," IEEE Photonics Journal 14 (2), pp. 1-11, (2022)
35. Clark, G., Raniwala, H., Koppa, M., Chen, K., Leenheer, A., Zimmermann, M., Dong, M., Li, L., Wen, Y.H., Dominguez, D. and Trusheim, M., 2023. Nanoelectromechanical control of spin-photon interfaces in a hybrid quantum

    system on chip. arXiv preprint arXiv:2308.07161.
36. Sohn, YI., Meesala, S., Pingault, B. et al. Controlling the coherence of a diamond spin qubit through its strain environment. Nat Commun 9, 2012 (2018).
37. Meesala, S., Sohn, Y.I., Pingault, B., Shao, L., Atikian, H.A., Holzgrafe, J., Gündoğan, M., Stavrakas, C., Sipahigil, A., Chia, C. and Evans, R., 2018. Strain engineering of the silicon-vacancy center in diamond. Physical Review B, 97(20), p.205444.
38. Guo, X., Stramma, A.M., Li, Z., Roth, W.G., Huang, B., Jin, Y., Parker, R.A., Martínez, J.A., Shofer, N., Michaels, C.P. and Purser, C.P., 2023. Microwave-based quantum control and coherence protection of tin-vacancy spin qubits in a strain-tuned diamond membrane heterostructure. arXiv preprint arXiv:2307.11916.
39. Assumpcao, D.R., Jin, C., Sutula, M., Ding, S.W., Pham, P., Knaut, C.M., Bhaskar, M.K., Panday, A., Day, A.M., Renaud, D. and Lukin, M.D., 2023. Deterministic Creation of Strained Color Centers in Nanostructures via High-Stress Thin Films. *arXiv preprint arXiv:2309.07935*.
40. Ruf, J.P., Paik, H., Schreiber, N.J. et al. Strain-stabilized superconductivity. Nat Commun 12, 59 (2021).
41. Zhengbao Yang, Shengxi Zhou, Jean Zu, Daniel Inman, High-Performance Piezoelectric Energy Harvesters and Their Applications, Joule,Volume 2, Issue 4,2018,Pages 642-697